# Big Issues for Big Data: challenges for critical spatial data analytics

Chris Brunsdon[1] and Alexis Comber[2]

[1] National Centre for Geocomputation, Maynooth University, Maynooth, Ireland

Email: Christopher.Brunsdon@mu.ie

[2] School of Geography, University of Leeds, Leeds, LS2 9JT, UK

Email: a.comber@leeds.ac.uk

**Abstract**

In this paper we consider some of the issues of working with big data and big spatial data and highlight the need for an open and critical framework. We focus on a set of challenges underlying the collection and analysis of big data. In particular, we consider 1) the issues related to inference when working with usually biased big data, challenging the assumed inferential superiority of data with observations, n, approaching N, the population ($n \rightarrow N$), and the need for data science analysis that answer questions of practical significance or with greater emphasis n the size of the effect, rather than the truth or falsehood of a statistical statement; 2) the need to accept messiness in your data and to document all operations undertaken on the data because of this support of openness and reproducibility paradigms; and 3) the need to explicitly seek to understand the causes of bias, messiness etc in the data and the inferential consequences of using such data in analyses, by adopting critical approaches to spatial data science. In particular we consider the need to place individual data science studies in a wider social and economic contexts, along the the role of inferential theory in the presence of big data, and issues relating to messiness and complexity in big data.

**Motivation**

Big data has, for a few years, been a dominant concept in many fields requiring empirical evidence. Terms such as *big data* and *data analytics* are replacing more traditional terms such as *data set* and *statistical analysis*, generally and in the area of applied quantitative geography. However, are there differences between these sets of terms that indicate something more complex than data with very large $n$ or $m$ values, implying a different practice to statistical analysis? Such questions are difficult to answer because there are no clear definitions of what constitutes a data set, a data analytics technique and so on. If one adopts the idea of a 'data cube', a three-dimensional array of location, time and variable - see for example Berry (1964) - several well established multivariate statistical analysis techniques could be applied to such data. One example is the *factor analysis* outlined in Clark, Davies, and Johnston (1974) which explicitly draw upon the data cube idea. The dates of these citations suggest that this kind of data, and these analytical methods are far from new. However, one could argue that as well as size, big data encompasses more kinds of data, more sophisticated data structures than the data cube and more kinds of analytical techniques have been introduced between those days and the current era. The widespread adoption of the paradigms introduced by *Geographical Information Systems* of the late 1980s and early 1990s (Brunsdon and Singleton 2015) extended data cubes to include the boundaries of regions, the paths of roads and rivers, and the point locations of simpler objects and provided tools for bulk processing these. Gimblett (2002) used agent-based models for geographical simulations of large datasets, suggesting that at least some of the current ideas associated with big data, sophisticated data models and data analytics are the result of a gradual evolution – a steadily evolving universe rather than a big bang. However, understanding how the big data/data science/data analytics universe that now exists functions on a day to day basis is essential, regardless of its origins. In particular we may ask how, as geographers, we can think about the implications.

**Hammer or anvil?**

…*'the hammer and the anvil', now always used with the implication that the anvil gets the worst of it. In real life it is always the anvil that breaks the hammer, never the other way about: a writer who stopped to think what he was saying would avoid perverting the original phrase.* - Orwell (1950)

One interesting aspect of data analytics is the prominence of spatial data. Whereas, within the statistics community spatial statistics may have been regarded as a niche area, in data analytics (and particularly marketing) dealing with locational data is seen as a core activity. As noted by Singleton and Arribas-Bel (2019):

*many contemporary "Big Data" are generated by companies whose activities are also mediated digitally, but often have clear spatial and geographical dimensions to their operations.*

It looks as though GI Scientists and geographers are starring in the show whether they planned to or not. However, as with data analytics and statistics, there is a danger that the emergent discipline ignores some important lessons learned from its predecessors. If GIScience has become a key player here, it is vital that it ensures such lessons are heeded. The *ecological fallacy* (Robinson 1950) will not go away simply because one uses machine learning algorithms proclaimed to be 'state of the art' on spatially aggregated data. Thus, quantitative geographers must simultaneously embrace and influence these trends: the theoretical community must participate in debates, and those involved in practice must lead by example. The remainder of this paper outlines a number of specific big data issues that we feel should be considered.

**Big Data**

There's always been big data - too big for a Commodore 64, too big for an IBM PC, too big for Excel, too big to fit on a hard drive. Workarounds have included software and hardware updates - eg new versions of Excel, bigger hard drives, cloud computing - as well as temporary fixes such as bespoke coding for specific problems. Addressing the technical issues of handling inconveniently large data sets is an evolving process, with what was 'big' a decade ago no longer considered as such. However, this is perhaps 'big data', but not 'Big Data'. More recently the term "Big Data" has taken on a broader meaning – reflecting not only technical issues, but a sea-change in data sources, data collection and approaches - with new analytical challenges but also wider social and cultural implications attracting much media attention. Cukier and Mayer-Schoenberger (2013) argue that the "Big Data" phenomenon is characterised by three things (as summarized by O'Neil and Schutt, 2014):

- Collecting and using a lot of data rather than small samples.
- Accepting messiness in your data.
- Giving up on knowing the causes.

The first of these has implications in terms of inference - to be discussed in the **Inference** section. The second will be addressed in the **Big Bad (messy) Data** section. The third - which we would regard as the most concerning - we address first in the **Critical Data Analysis** section.

**Critical Data Analysis**

A critical data analysis is one in which the practitioners are aware of both the limitations of the data and the way that analysis results and models are deployed. As well as considering techniques of data analysis, many argue (eg Kitchin, 2014) for critical reflection on the assumptions often adopted by the media and technical literature, that challenge the notion of big data as objective, and all-encapsulating (Iliadis and Russo 2016). This has led to the field of *critical data studies* (Dalton and Thatcher, 2014) which argues for the need "*to explore the ways in which [data] are never simply neutral, objective, independent, raw representations of the world, but are situated, contingent, relational, contextual*" (Kitchin and Lauriault, 2014, p5). It emphasises the need to think about how data analyses are deployed - the cooking of data in collection and analysis - and to consider '*the technological, political, social and economic apparatuses and elements that constitutes and frames the generation, circulation and deployment of data*' (Iliadis and Russo, 2016, p2). We should '*think about big data science in terms of the common good and social contexts*' (Iliadis and Russo, 2016, p5). This criticality can be extended to embrace critical views of the more technical aspects of the data analysis. Quoting O'Neil and Schutt (2014, p352)

*'We'd like to encourage the next-gen data scientists to become problem solvers and question askers, to think deeply about appropriate design and process, and to use data responsibly and make the world better, not worse.'*

And they advocate taking on board Derman's (2011) *Hippocratic Oath of Modeling*, in particular the declaration *'I understand that my work may have enormous effects on society and the economy, many of them beyond my comprehension'*

One could easily add 'the environment' to 'society and the economy' in that statement. A typical concern may then be the wider implications of misinterpreting the inferential aspects of a particular analysis. Returning to O'Neil and Schutt (2014, p354) *'Even if you are honestly skeptical of your model, there is always the chance that it will be used the wrong way in spite of your warnings'*.

This presents a number of interlinked challenges to society as ever-growing volumes of data are collected and mashed together and as some form of data analysis is the de facto way of providing evidence in support of some policy or decision. There is a need to be aware of the inherent biases in data that were *not* collected under some form of experimental as well as the assumptions in the technological, political, social and economic "data assemblages" within which big data are deployed (Kitchin and Lauriault, 2014). The former suggests the need to explicitly account for and challenge the assumption that '*the "we" of those who emit data is a statistically representative "we"* ' (Dalton et al, 2016, p4) and to acknowledge biases in data origins and coverage (more "*data fumes*" are emitted in the global North (Dalton et al, 2016). The deployment of big data is often framed in philosophies thar are grounded in "letting the data speak" under "a veil of openness and transparency and responsible data practices" (Iliadis and Russo, 2016, p3). Additionally, there is perhaps a concern that critical data theorists rarely 'crunch numbers' but reciprocally, number crunchers rarely consider critical data analysis. An awareness of work such as the above may help to provide a more grounded approach to data science.

**Inference**

A key task of applied statistics / data analytics is to connect data collected to the estimation of some quantity or the validation of a hypothesis. Put more simply, does the data support a given theory? The *null hypothesis significance test* (NHST) (Gigerenzer 2004) accommodates both classic inference (Fisher, 1925) and the possibility of alternative hypotheses (Neyman and Pearson, 1928), despite their epistemological differences Perezgonzalez (2014; 2015). Despite its vague conceptualisation the NHST is adopted as though it were a gold standard in many scientific disciplines. However the NHST requires careful, informed and considered formulation for the results to be of practical use, with well reported limitations when applied to big data. There are alternatives including Bayesian approaches, the use of data visualisation (especially in geographical applications, where spatial pattern is of primary importance), and exploratory approaches which may provide useful, if more informal, inferential tools.

The above considerations are not specific to Big Data - however, they certainly do apply when statistical tests are applied to large data sets - and so deserve attention in this context. A further big data problem relates to Cukier and Mayer-Schoenberger's idea and assumption of inferential superiority of *n=all* over sampled data, as the number of observations, n, approaches N, the population ($n \rightarrow N$).

To illustrate this, assume that house price data for every house sold in the UK in a given year are available with copious attributes are available. Suppose the aim was to test a null hypothesis that the kind of letter box influenced house price and that there were two kinds: up-down and left-right. One can simply split the data in two and compute the average price for each group. Suppose on doing this, we find that there is a £25 difference in favour of left-right. Were we to carry out a significance test, it would almost certainly reject an $H_0$ of no difference ($n$ would be huge and the standard error of the estimated difference tiny). However, this difference is of no consequence due to a poorly formulated hypothesis. Perhaps the question here should have been 'is the difference greater than £50?' or some other quantity deemed to be of *practical* significance. Or emphasis could be placed on the size of the effect, rather than the truth or falsehood of a very specific statement. If this is done, it is context, rather than a numerical procedure, that determines what is 'significant'.

Another problem with the *n=all* idea is that in reality, although $n$ is large, it isn't 'all' and the discrepancy is problematic. Sample-based statistical procedures make use of *random* samples (each member of the population is equally likely to be sampled) or *stratified sampling* (so that sampling is independent of the effect being measured).

However the messiness of Big Data arising from the way in which it is acquired is often problematic as the mode of collection is not documented. Thus, although $n$ may be very large (possible, say 80% of the population), it is hard to guarantee that the missing 20% is representative of the population rather some distinct subgroup (who do not use social media, lack a smartphone, a loyalty card etc) such that the 80% is a biased sample (albeit a very large one). This implies that statistics such as the sample mean, or a correlation between a pair of variables, will also be biased. Meng (2016) considers this quantitatively for the estimation of sample means and proportions, by considering the correlation between whether observation $i$ is included in the sample (the binary variable $w_i$) and $x_i$ - the observed value whose mean is to be computed - $\rho$. He uses this to address a motivating question: 'Which one should I trust more: a 1% survey with 60% response rate or a self-reported administrative dataset covering 80% of the population?'. He finds that $|\rho| < 0.0034$ in order to trust the latter. Thus, with even quite slight bias, the administrative big dataset proves to be the worst option for this particular task.

Thus there are a number of inferential challenges when working with big data. The first is that hypotheses need to be carefully formulated and tested to generate inferences of practical significance. The second is that, it is difficult to exchange quality for quantity, suggesting that an unquestioning belief that $n$ being very large makes hypothesis testing infallible could lead to some problematic decision making and policy implementation. Both of these issues suggest that one must look closely at the idea of statistical testing and inference when working with Big Data - it seems that even if one is willing to accept that standard techniques may work reasonably well when data sets are smaller, their largeness pushes their utility to the limit (or possibly beyond).

**Big Bad (messy) Data:**

A third characteristic of Big Data is its messiness. This can manifest itself in bias from undesigned data collection (as above), but there are other ways in which data can be messy. Some common examples of messiness in big data are:

- Erroneous or missing information recorded.
- Complicated data formats (data not a data frame).
- Awkward data formats (often due to poorly designed proprietary formats).

These issues frequently occur in traditional data analysis, as well as when working with Big Data.

*Erroneous or missing data* will lead to difficulties for any kind of analysis. These can arise transcription errors or some malfunction of an automated recording system with varying severity for downstream analysis depending on the nature of the error. While manual transposition errors are more common with "small" data (Noone et al 2017 and Ryan et al, 2018) and not a practical consideration for Big Data, missing data are. Typically information can be missing due to non-response to a question in a survey, or in the case of automated data collection, when a data recording instrument fails. In terms of impact, this is similar to biased sampling as discussed earlier: if the chance of an observation being missing is not independent of the outcome, then simply omitting the missing data in an analysis could lead to a biased analysis. For example if a temperature sensor measuring temperature at noon malfunctions over a number of days in July in the UK, this would very likely lead to an underestimate of mean annual temperature at noon. In this case, the missing data can be simulated by noting the correlation between consecutive daily temperatures and simulating several series of observations that match the last observation before breakdown, and the first after the sensor began to function again. The missing information can be estimated by averaging the simulations (grouped by day) and even gain some insight into the uncertainty of these estimations by computing standard deviations. However, it is important to flag the estimated values rather than treat them as though they were direct measurements. One reason for this is that in the future, someone analysing the data may fit a similar model to the one used to estimate the missing observations - giving rise to an over-optimistic belief in the reliability of that model.

The phrase *complicated data formats* describe formats that differ notably from the 'data frame' model containing a number of rows (cases or observation) with a fixed number of variables in columns of the same type (i.e. integer, character variable and so on). In some cases the complication may be fairly minor, for example where column delimiters in the contents of cells confuse the reading of the data. This can be overcome with some judicial editing, search and replacing, column delimiter counting etc applied line by line to the raw data prior to reading the data frame. All data are spatial – they are collected some*where* and a further format complication relates to the way that geographical information is stored,. Typical 'vector' formats for geographical data record objects as collections of *points*, *lines* or *polygons*, with the last two storing the coordinates of geographical features in a 'join the dots' style sequence of varying length, depending on the shape of the feature. A further complication is that there are a many possible map projections that could be used to encode location in the coordinates. A great deal of Big Data involves processing this textual data, such as the content of tweets. This is perhaps the most 'complex' in our definition as a great deal of valuable information is contained in the tweet. Tweets can be represented in data frame format (for example with columns such as 'Tweeter', 'Time of Tweet', 'Content of Tweet'), helped by the use of hashtags (#) and @ symbols to identify who engage in dialogs, and subject matter. The tweet content is more complex and although there are ways of dealing with this in a standard data frame it requires large amount of pre-processing and data transformation is required (Silge and Robinson, 2017). Similar considerations apply to working with web data where tools can be used to carefully extract information from web sites before reorganising at least part of the data into a data frame. Finally, graph or network data is another frequently encountered data format. This can be represented as a series of data frames, with one frame containing columns for edge ID, from-node ID and to-node ID and others containing node and edge attributes, using the IDs as linking keys. Arguably the graph or network is rapidly becoming another standard data structure for Big Data, as it is able to encode relationships between records and features in different ways, can be used to represent geographic features and their topological relationships.

In contrast to complicated data formats which are well defined, *awkward data formats* may be mis-specified, partially specified or not well designed. The problems manifest themselves in inappropriate structures, lack of structure or more technical issues. An example of a technical issue may be Excel's inconsistent date storage format between Windows-based systems and Apple Macs (Mott 2013). As Woo (n.d.) observes '*As it turns out, Excel 'supports' two different date systems: one beginning in 1900 and one beginning in 1904. Excel stores all dates as floating point numbers representing the number of days since a given start date, and Excel for Windows and Mac have different default start dates (January 1, 1900 vs. January 1, 1904). Furthermore, the 1900 date system purposely erroneously assumes that 1900 was a leap year to ensure compatibility with a bug in - wait for it - Lotus 123. You can't make this stuff up.*' This results in a 4-year difference in recorded dates between the two systems. It also adds a new dimension to drives towards reproducibility - not only are the data and code required to be made open, but also information about the operating systems involved. On occasion, code must then be used to correct for this. Another technical issue relating to dates is that Excel sometimes autoformats numbers not intended to be dates (such as serial numbers) as dates. Some character strings may also be 'automatically' and inappropriately converted to dates such as the gene sequence number 'Oct-4' (Broman and Woo 2018). The problem with Excel is that it tries to be more than a data frame, allowing the user to add notes against particular data rows, include graphics, highlight cells, etc. Making notes about particular observations is good practice, but should be done in a separate file: attempting to mix metadata, visualisation and analysis generally impedes the data sharing process, and introduces the possibility of errors being introduced without being noticed. Certainly, one practical requirement is that any Excel spreadsheet intended to share information should be saved in csv format, without any loss or distortion of the original information.

Big data is messy and presents a number of challenges for its correct and appropriate incorporation into analyses. Most of these relate to different kinds of cleaning, pre-processing, tidying and remedial work on the data structure to get the data into a flat data frame format. Complicated data formats (as viewed here) are not in the standard data frame format but are consistently defined in raw form and can generally be analysed by providing some code to process the raw data. Awkward data formats should be avoided as they require often bespoke pre-processing should extract data and data files should just contain flat data tables, metadate notes and guidance should be linked to but separated from the data. Text and web data need some careful reprocessing to render them into formats suitable for data analysis, missing data can be infilled (e.g. through simulation) but must be flagged as such, geographic big data should be accompanied with metadata of the map projection being used as a minimum. All of these operations should be documented, highlighting the need for openness and reproducibility paradigms to be adopted especially for geographic big data (Brunsdon and Comber, 2020).

**Closing Comments**

The above discussion has considered a number of challenges and related rubrics that we feel should be kept in mind when working with big data or data more generally, and data analytics. These include:

- an *a priori* expectation of biases to be inherent in all data that were not collected under some formal experimental design (i.e. challenge the assumption of representativeness).

- *de facto* questioning of the assemblages within which data are deployed (including your own) – they will not be objective or neutral, despite their veil of transparency (i.e. be a *critical* spatial data scientist).

- not simply letting the data speak without careful examination of the inferential pedigree of the analysis that has resulted in the message (i.e. hypotheses still need to be carefully formulated).

- not being blind to the vast potential for inferior inference as the number of observations, n, approaches N, the population ($n \rightarrow N$), and the potential for superior inference from a small set of highly targeted observations or samples.

- having an expectation of messy data and being prepared to spend a lot of time munging and wrangling data.

The orgins of these challenges relate to changes in the way that data are created and it is perhaps here that data science, or data analytics, changes the focus. Historically, early statisticians developed their set of practices in an era when there were no computers, and data were often recorded manually, so that many of the practical issues identified in the preceding did not exist. The final section above proposed 'good practice' approaches to sharing data and reproducibility, but in many cases data are not created in this way, and discrepancies must be dealt with. On occasion this may involve detective work - essentially recreating details of the data formats that are vaguely specified, not specified, or mis-specified - many case studies are given in McCallum (2013). This too is a part of data science. Such challenges relate to all data secondary data but have greater salience in the context of increased availability, volumes and diversity of data, all with some form of location attached. They apply to any data not collected by you and / or not under some form of experimental design.